# Optoelectronic and Thermoelectric Properties of High-Performance AlSb Semiconductors

Dilshod Nematov [a,b*], Amondulloi Burkhonzoda [a**], Iskandar Raufov [a], Sherali Murodzoda [a], Saidjafar Murodzoda [a], Sakhidod Sattorzoda [a], Anushervon Ashurov [a], Makhsud Barot Islomzoda [c], Kholmirzo Kholmurodov [a]

[a] S.U. Umarov Physical-Technical Institute of NAST, Dushanbe 734063, Tajikistan
[b] School of Optoelectronic Engineering & CQUPT-BUL Innovation Institute, Chongqing University of Posts and Telecommunications, Chongqing 400065, China
[c] Tajik National University, Dushanbe 734025, Tajikistan
 Corresponding author's E-mail: *dilnem@mail.ru (D. Nematov), ** amondullo.burkhonzoda@mail.ru (A. Burkhonzoda)

**ABSTRACT**

This study presents a comprehensive first-principles investigation of the optoelectronic and thermoelectric properties of aluminum antimonide (AlSb) in its cubic (F-43m) and hexagonal (P6$_3$mc) phases. Structural optimization was performed using the SCAN functional, and all electronic and optical properties were evaluated using the modified Becke-Johnson potential combined with the Hubbard correction (mBJ+U), which best describes the band-edge electronic structure, explicitly accounting for the contribution of the d-states of the Sb half-core, which cannot be adequately accounted for by conventional functionals and may be overestimated by hybrid approaches. Both AlSb phases are found to be quasi-direct bandgap semiconductors, with calculated band gaps of 1.71 eV for the cubic phase and 1.50 eV for the hexagonal phase, in good agreement with available experimental data. The optical response reveals strong absorption in the visible and ultraviolet regions, moderate reflectivity, and high refractive indices, indicating pronounced light-matter interaction characteristic of III-V semiconductors. The hexagonal phase exhibits enhanced low-energy optical absorption due to its reduced symmetry and narrower band gap. Thermoelectric analysis demonstrates large negative Seebeck coefficients, thermally activated carrier generation, and a monotonic increase of the power factor with carrier concentration for both phases. The cubic phase shows higher power factor values due to enhanced carrier mobility, whereas the hexagonal phase benefits from reduced thermal conductivity, which is favorable for thermoelectric performance at elevated temperatures. These results establish AlSb as a multifunctional semiconductor with tunable optoelectronic and thermoelectric properties and highlight the importance of an accurate treatment of Sb d-electron effects for reliable property prediction.

# 1. INTRODUCTION

The development of modern optoelectronic and thermoelectric devices depends on materials that can simultaneously maintain stable electrical transport, withstand thermal loads, and interact efficiently with light. Traditional semiconductors such as Si, GaAs, CdTe and halide perovskites have held a leading position in photovoltaic and infrared technologies for many years [1-4]. At the same time, their performance is often limited by temperature-induced degradation, chemical instability, or other practical issues that restrict their use in more demanding conditions [5-14]. These limitations have encouraged the search for compounds whose properties can be tuned more flexibly and which offer greater structural and thermal robustness.Another trend that has become increasingly important is the use of multifunctional materials that combine optical absorption, charge transport, and thermoelectric conversion within a single system. Such an approach is attractive because it reduces the number of layers in a device and simplifies fabrication while maintaining high efficiency [15-17]. Achieving this balance, however, requires a semiconductor with suitable band-edge dispersion, strong bonding stability and predictable phonon behavior.

Aluminum antimonide (AlSb), a III-V compound with a moderate direct band gap of about 1.6 eV, has drawn renewed attention for these reasons. It exhibits relatively high carrier mobility and is compatible with established epitaxial growth techniques [8-10], which makes it attractive for infrared optoelectronics, photovoltaic cells, radiation detection, and thermoelectric applications. Although the cubic zinc-blende phase (F-43m) is the stable form at ambient pressure, a hexagonal $P6_3mc$ modification can be stabilized under epitaxial strain or in nanostructured forms [11-13]. Because the symmetry and bonding patterns differ in the two structures, their optical absorption, band gaps, and transport properties also differ, creating opportunities for targeted property tuning.

Previous studies have shown that the thermoelectric performance of bulk AlSb has been limited by relatively high lattice thermal conductivity and modest Seebeck coefficients [14]. Nonetheless, recent theoretical and experimental works suggest that phonon engineering, nanostructuring, and alloying can substantially enhance the thermoelectric figure of merit [15-17]. Furthermore, the development of first-principles methods, such as the modified Becke-Johnson (mBJ) potential and hybrid HSE06 functional, allows accurate prediction of optical transitions, carrier transport, and phonon-limited heat conduction, making it possible to reassess the intrinsic optoelectronic potential of AlSb.

Recent investigations of low-dimensional and pressure-modified AlSb systems have revealed a direct band gap in the range of 1.5-1.7 eV, high absorption coefficients, and strong polarizability comparable to that of other III-V materials like GaSb and InSb [23-27]. However, comprehensive comparative studies of the cubic and hexagonal phases, particularly addressing their optoelectronic and thermoelectric behavior, remain limited. Understanding how crystal symmetry and electronic configuration influence photon absorption, carrier mobility, and thermal transport is essential for the rational design of AlSb-based energy conversion materials [28-30]. Recent studies also highlight an increasing interest in III-V materials that can operate as multifunctional platforms, serving simultaneously as optical absorbers, charge-transport layers, and thermoelectric converters [15-17, 23-27]. AlSb is particularly attractive in this context due to its combination of a moderate direct band gap, strong light-matter interaction, and promising Seebeck response. Integrating optoelectronic and thermoelectric functionalities within a single semiconductor platform can significantly simplify device architectures and improve energy-conversion efficiency, especially in high-temperature or infrared-operating environments [8-10, 15-17].

Previous first-principles studies on AlSb have mainly focused either on the electronic and optical properties of the cubic F-43m phase or on pressure-induced transitions to dense B1/B2 structures, typically considering a limited subset of physical properties and a single crystal structure [1,2,13,21,24,27,29]. In addition, most thermoelectric reports treat bulk AlSb within simplified transport models or without explicitly addressing the role of phonon stability, elastic anisotropy and phase competition between the cubic and hexagonal polymorphs. As a result, there is still no unified picture that consistently links structural, vibrational, electronic, optical and thermoelectric behavior of AlSb in both phases.

In this work, we fill this gap by performing a comprehensive first-principles investigation of AlSb in its cubic F-43m and hexagonal P6$_3$mc phases. For the first time, we combine SCAN-relaxed geometries with mBJ and HSE06 electronic-structure calculations and Boltzmann transport modeling to obtain a consistent description of phase stability, phonon dynamics, optoelectronic behavior and thermoelectric performance in both polymorphs. This combined approach allows us to clarify how crystal symmetry and bonding anisotropy control the band gap, optical absorption, lattice stiffness, thermal transport, and Seebeck response under ambient condition.
.

## 2. COMPUTATIONAL DETAILS

The optoelectronic and thermoelectric properties of cubic (F-43m) and hexagonal (P6$_3$mc) aluminum antimonide (AlSb) were investigated using density functional theory as implemented in the Vienna *ab initio* simulation package (VASP) [31]. The interaction between electrons and ions was described using the projector augmented-wave (PAW) method, with valence electron configurations Al: $3s^23p^1$ and Sb: $4d^{10}5s^25p^3$. Structural optimization of both phases was performed using the SCAN meta-GGA exchange-correlation functional [32]. Convergence tests were carried out to ensure numerical accuracy, resulting in a plane-wave cutoff energy of 700 eV and Γ-centered k-point meshes of 7×7×7 for the cubic phase and 12×12×5 for the hexagonal phase. Following full structural optimization, all optoelectronic properties were calculated using the modified Becke-Johnson potential combined with a Hubbard correction (mBJ+U) [33, 34]. This approach was chosen to overcome the well-known band gap underestimation of local and semi-local functionals (LDA, GGA, SCAN), which is particularly pronounced in AlSb due to the hybridization of Sb 5p states with semicore d electrons. The Hubbard U correction [34] was applied to the Sb d orbitals in order to properly account for their localized character and to reduce self-interaction errors, leading to a more accurate description of the band-edge electronic structure. Electronic band structures, total and partial densities of states (DOS and PDOS), and optical properties, including the complex dielectric function, absorption coefficient, refractive index, reflectivity, and energy-loss function, were calculated within the linear response formalism based on the mBJ+U electronic structure. Thermoelectric transport coefficients were evaluated using the BoltzTraP2 code interfaced with Quantum ESPRESSO 7.4.1 [35]. Calculations were performed within the constant relaxation time approximation, employing a dense k-point mesh (35×35×35) to ensure convergence of transport integrals.

## 3. RESULTS AND DISCUSSION

Aluminum antimonide crystallizes in two structural forms: a cubic lattice (space group F-43m) and a hexagonal lattice (space group P6$_3$mc). Before analyzing its properties, convergence tests were performed to determine optimal parameters for stable energy minimization. The effects of plane-wave cutoff energy and k-point mesh density on total energy convergence were examined using Γ-centered and Monkhorst-Pack schemes. As shown in Figure S1, the total energy stabilized at about 700 eV, and further increases produced negligible changes while raising computational cost. Both sampling methods demonstrated consistent convergence, confirming the accuracy of the chosen Brillouin zone discretization and the reliability of the calculations.

For both AlSb phases, k-point optimization was carried out at ENCUT = 315 eV (approximately 1.3 × ENMAX). The optimal meshes were 7×7×7 (Γ-centered) for the cubic and 12×12×5 for the hexagonal structure (Figure S2). These parameters were used in all electronic-structure and optical-property calculations. The Monkhorst-Pack grid provided uniform sampling for highly symmetric systems, whereas the Γ-centered mesh ensured accurate representation of states near the Γ point, which is important for lower-symmetry crystals. The difference in total energy between the two schemes was below 1 meV per atom, confirming sufficient numerical precision.

Following convergence validation, full structural optimization was performed for both phases. Equilibrium lattice constants, cell volumes, and interaxial angles were evaluated using several exchange-correlation functionals and compared with experimental data (Table S1). The PBE functional, part of the generalized gradient approximation (GGA) family, predicts the largest lattice constants, while LDA yields the smallest. PBEsol and SCAN values lie between these limits and closely match experimental results [37,38]. For example, for the cubic phase, lattice constants are a = 6.233 Å (PBE), 6.169 Å (PBEsol), 6.121 Å (LDA), and 6.172 Å (SCAN). The trend arises from intrinsic differences in the treatment of electron correlation: PBE tends to overestimate volumes, LDA underestimates them, while SCAN provides reliable predictions for both structural and energetic parameters.

To assess the relative thermodynamic stability of the cubic (F-43m) and hexagonal (P6$_3$mc) phases of AlSb, the difference in free energy (ΔF) was evaluated as a function of temperature, as shown in Figure 1. At low temperatures, the calculated free-energy difference is negative, indicating that the cubic F-43m phase is thermodynamically more stable than the hexagonal P6$_3$mc phase. This result is consistent with experimental observations, where AlSb predominantly crystallizes in the zinc-blende structure under ambient conditions. The negative value of ΔF reflects the lower total energy of the cubic phase, arising from its higher symmetry and more favorable bonding configuration. With increasing temperature, ΔF increases monotonically and approaches zero, reflecting a gradual reduction of the free-energy difference between the two phases. This behavior originates from entropic contributions to the free energy, particularly vibrational entropy, which is enhanced in the lower-symmetry hexagonal structure. However, the free-energy curves do not intersect, and no temperature-induced thermodynamic phase transition is predicted within the investigated temperature range. At elevated temperatures, the reduced magnitude of ΔF indicates that the hexagonal P6$_3$mc phase becomes progressively more competitive in terms of thermodynamic stability, although the cubic phase remains energetically favored. This trend suggests that while the cubic structure represents the equilibrium ground state, the hexagonal phase should be regarded as a metastable modification that may be stabilized under non-equilibrium synthesis conditions, such as thin-film growth, strain engineering, or high-temperature processing.

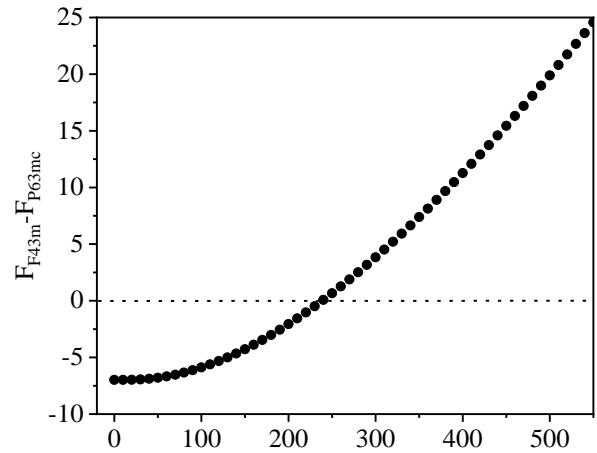

**Figure 1.** Temperature dependence of the free-energy difference between cubic and hexagonal phases of AlSb.

Although the differences between the two AlSb phases are minor, they become more evident at higher temperatures, revealing distinct levels of thermodynamic stability. The calculated formation energies, -1.316 eV for the cubic F-43m and -1.258 eV for the hexagonal P6$_3$mc phase, confirm the superior thermodynamic stability of the cubic form. These findings are consistent with the thermodynamic results and provide a sound basis for further modeling, including phase diagram development and evaluation of stability in solid-state systems and heterostructures.

Following full structural optimization, the electronic band gaps of cubic and hexagonal AlSb were calculated using the mBJ+U approach. Conventional LDA, GGA, and SCAN functionals systematically underestimate the band gap due to self-interaction errors, which are enhanced in AlSb by the hybridization of Sb 5p states with semicore d electrons, while hybrid functionals such as HSE06 may overestimate the gap and significantly increase computational cost. The mBJ+U method accounts for the localized character of Sb d electrons and provides an improved description of band-edge states without excessive gap opening. The band gap values were determined from the electronic band structures as the energy difference between the conduction band minimum and the valence band maximum. The total density of states (DOS) confirms the obtained gaps by exhibiting a clear energy region with zero DOS around the Fermi level. For the cubic F-43m phase, the band gap is about 1.7 eV, while for the hexagonal P63mc phase it decreases to approximately 1.5 eV. The resulting band gap values are summarized in Table 1 and compared with available experimental data.

**Table 1.** Calculated band gap energies of cubic and hexagonal AlSb obtained using the mBJ+U approach, compared with available experimental data.

| AlSb | Calculation (eV) | Experimental (eV) |
|:---:|:---:|:---:|
| F43m | 1.71 | 1.63[39], 1.75[40], 1.81[41] |
| P6$_3$mc | 1.50 | - |

The results show that mBJ+U produce band gap values in close agreement with each other and consistent with experimental data. For the cubic F-43m phase, the calculated band gap is 1.78 eV, matching well with the experimental values of 1.63-1.81 eV [39-41]. This agreement demonstrates the reliability of the computational approach and the accurate treatment of exchange-correlation effects in the AlSb system [42-53]. The hexagonal P6$_3$mc phase exhibits a slightly narrower band gap, which suggests stronger overlap between the valence and conduction bands and potentially higher carrier mobility. The reduced symmetry of the P6$_3$mc phase originates from its different stacking sequence and non-equivalent Wyckoff positions of Al and Sb, which cause small distortions in Al-Sb bond lengths and bond angles. These structural deviations from the ideal tetrahedral environment of the cubic phase weaken the sp$^3$ hybridization and increase the density of electronic states near the band edges, resulting in the observed narrowing of the band gap. The narrower band gap (Eg) of the hexagonal phase arises from its lower lattice symmetry and altered local coordination, which produce a denser distribution of electronic states near the band edges and reduce the strength of Al-Sb s-p hybridization. This effect shifts the conduction-band minimum downward and results in a smaller band gap compared to the cubic phase.

Next, using carefully relaxed structures obtained with the SCAN functional, we investigated the electronic properties of AlSb. The band gap widths were refined using the mBJ+U approach, while the electronic structure was analyzed in detail through the calculated band structures and total density of states (Figures 2 and 3), as well as the partial density of states (Figure 4). The electronic band structure and total DOS of cubic AlSb (F-43m) calculated using the mBJ+U approach are shown in Figure 2. The cubic phase exhibits a direct band gap with well-defined band edges, consistent with the value reported in Table 1. The pronounced dispersion near the conduction-band minimum indicates favorable carrier transport.

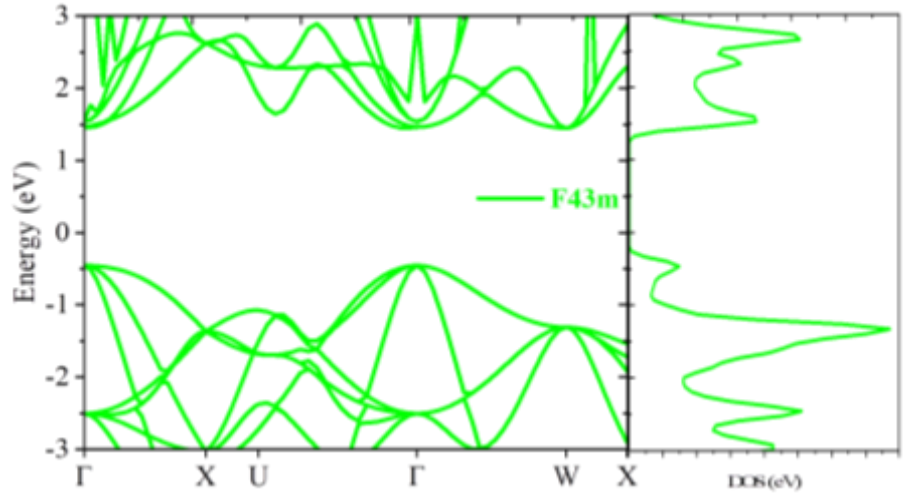

**Figure 2.** Electronic band structure and total density of states (DOS) of cubic AlSb (F-43m) calculated using the mBJ+U method.

The electronic band structure and DOS of hexagonal AlSb (P6$_3$mc) are presented in Figure 3. Similar to the cubic phase, the hexagonal structure exhibits a direct band gap; however, the gap is reduced, in agreement with Table 1. The band dispersion near the valence-band maximum is slightly flatter, leading to an increased density of states near the band edges.

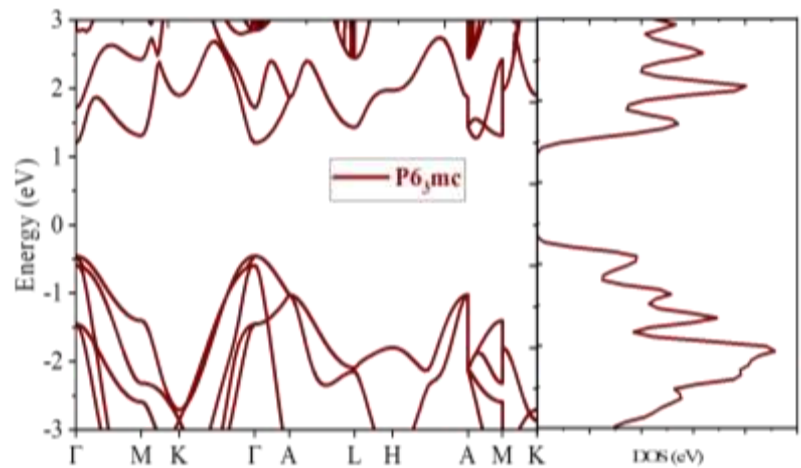

**Figure 3.** Electronic band structure and total density of states (DOS) of hexagonal AlSb (P6$_3$mc) calculated using the mBJ+U method.

Compared to the cubic phase, the hexagonal structure shows flatter band dispersion near the valence-band maximum, which results in an increased density of states close to the band edges. This behavior originates from the different stacking sequence and the presence of non-equivalent Wyckoff positions for Al and Sb atoms, leading to distortions in Al-Sb bond lengths and bond angles. These structural deviations weaken the ideal sp$^3$ hybridization found in the cubic phase, enhance the electronic-state density near the Fermi level, and shift the conduction-band minimum to lower energies, resulting in band-gap narrowing.

Additional insight is provided by the partial density of states shown in Figure 4(a). The valence band is dominated by Sb *p* states with a smaller contribution from Al *p* orbitals, while the conduction band is primarily composed of Al *s* states with minor Sb *s* character. This confirms strong *s-p* hybridization at the band edges, typical for III-V semiconductors. The Sb *d* states are located far from the Fermi level and show negligible contribution near the band edges, indicating that their influence on the band gap is indirect. The corresponding PDOS shown in Figure 4(b) reveals that the valence band remains dominated by Sb *p* states, while the conduction band is mainly formed by Al *s* states. Compared to the cubic phase, the hexagonal structure exhibits a higher PDOS near both the valence- and conduction-band edges. This enhancement originates from the reduced crystal symmetry and altered local coordination of Al and Sb atoms, which lead to distortions in Al-Sb bond lengths and bond angles. These distortions weaken the ideal $sp^3$ hybridization observed in the cubic phase and result in a denser distribution of electronic states near the band edges, thereby contributing to the observed band-gap narrowing.

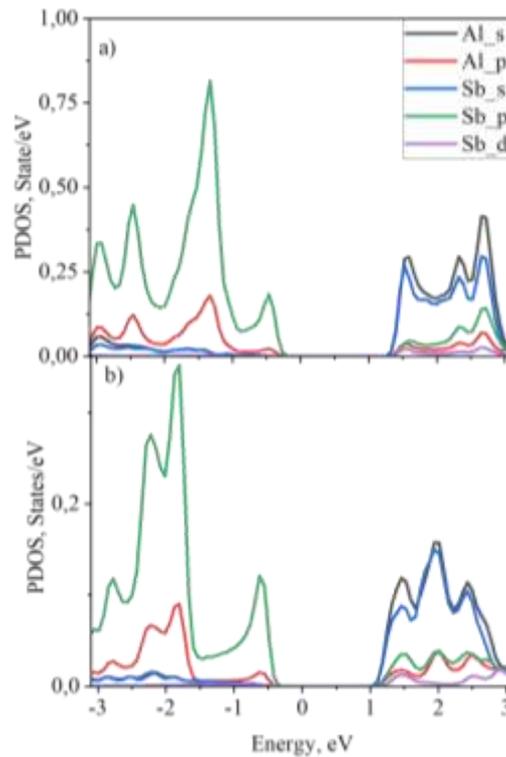

**Figure 4.** Partial density of states (PDOS) of AlSb calculated using the mBJ+U method: (a) cubic F-43m phase and (b) hexagonal P6$_3$mc phase, showing the orbital contributions of Al (*s*, *p*) and Sb (*s*, *p*, *d*) states.

Both of the AlSb phases are direct-gap semiconductors with band edges governed by Sb *p* and Al *s* states. The cubic F-43m phase shows a wider band gap and lower PDOS near the band edges, consistent with its higher symmetry and stronger $sp^3$ hybridization. In contrast, the hexagonal P6$_3$mc phase exhibits enhanced PDOS near the band edges and a reduced band gap, which may favor optical transitions and charge-carrier activity in lower-gap optoelectronic applications. The PDOS results are in excellent agreement with the calculated band structures shown in Figures 2 and 3, confirming consistent electronic trends across both phases.

To further investigate the optical response of AlSb, the frequency-dependent complex dielectric functions was calculated for the F-43m and P6$_3$mc phases using the mBJ+U electronic structure. Based on the dielectric function, key optical parameters including the absorption coefficient (α), extinction coefficient (k), energy-loss function (L), refractive index (n), and reflectivity (R) were derived. Figure 5 (a,b) shows the real ε1(ω) and imaginary ε2(ω) parts of the dielectric function for both AlSb phases. The onset of ε2(ω) occurs at photon energies consistent with the calculated band gaps, confirming that the optical absorption is governed by direct interband transitions. The cubic phase exhibits a slightly higher peak intensity in ε2(ω), reflecting stronger optical transition probabilities associated with its higher symmetry and more delocalized electronic states. The real part ε1(ω) displays normal dispersive behavior, with positive values at low energies and sign reversal at higher photon energies, indicating plasma resonance effects.

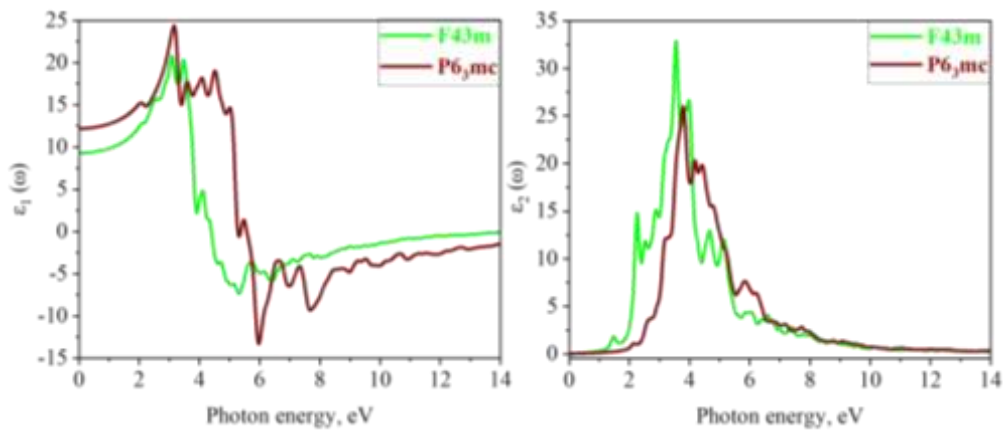

**Figure 5.** Real (a) and imaginary (b) parts of the dielectric function of F-43m and P6$_3$mc phases AlSb calculated using the mBJ+U method

The stronger maxima observed for the hexagonal phase indicate a higher density of electronic states near the Fermi level, which enhances optical transition probabilities and explains its stronger high-energy response. The extinction coefficient (Figure 6) follows the same general pattern as the absorption spectrum, reaching its maximum between 4 and 6 eV and gradually decreasing at higher photon energies, reflecting the dispersive nature of the optical response.

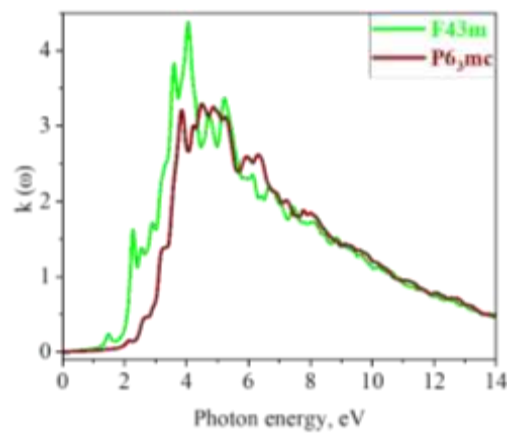

Figure 6. Energy-dependent extinction coefficient of cubic and hexagonal AlSb calculated using the mBJ+U method.

The absorption coefficients of both phases reach values on the order of $10^4$ cm$^{-1}$, typical of efficient light-harvesting semiconductors, confirming the potential of AlSb for thin-film photovoltaic applications (Figure 7). In the 2-12 eV range, multiple resonance peaks associated with interband transitions are evident. The energy loss function (Figure 8) displays a dominant plasmon resonance around 13 eV for both structures, corresponding to collective oscillations of conduction electrons and indicating similar plasma frequencies. Both AlSb phases demonstrate strong ultraviolet absorption and high transparency in the visible spectrum, confirming their suitability for optoelectronic and photonic devices such as photodetectors, light-emitting diodes, and solar absorbers.

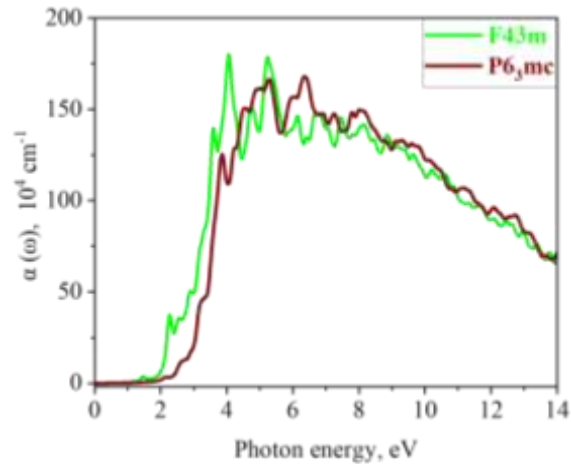

Figure 7. Energy-dependent absorption coefficient of cubic and hexagonal AlSb calculated using the mBJ+U method.

Multiple peaks in the 3-7 eV range originate from interband transitions between Sb *p* states in the valence band and Al *s/p* states in the conduction band, in agreement with the PDOS analysis. The hexagonal P6$_3$mc phase shows a slightly red-shifted absorption onset and enhanced absorption intensity at lower photon energies, reflecting its narrower band gap and higher density of states near the band edges. The extinction coefficient follows a similar trend, with pronounced maxima in the same energy range, confirming the strong dispersive optical response of both structures.

The calculated energy-loss function, shown in Figure 8, exhibits a pronounced plasmon resonance at high photon energies for both phases. This peak corresponds to collective oscillations of valence electrons and indicates similar plasma frequencies for cubic and hexagonal AlSb. The comparable plasmon energies suggest that the overall free-electron density and screening behavior are not strongly affected by the structural phase transition. These values are comparable to those reported for other III-V semiconductors such as InSb and GaSb, reflecting the high polarizability and strong covalent bonding characteristic of the Al-Sb system. The large refractive index indicates strong light-matter interaction, which enhances optical confinement in nanoscale devices. The reflectivity spectra show moderate values of up to 60%, suggesting efficient coupling of incident light into the material without excessive surface reflection. This property is advantageous for photovoltaic and optoelectronic devices where both absorption and internal photon management are critical.

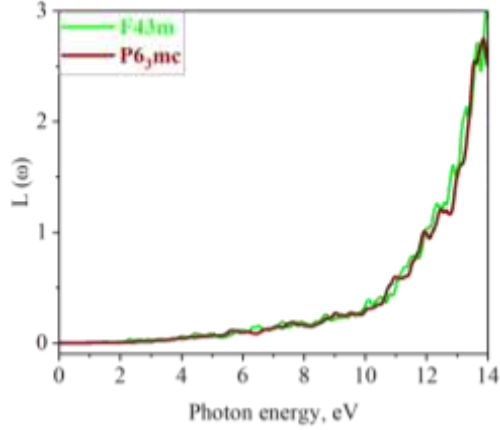

**Figure 8.** Energy-loss function of cubic and hexagonal AlSb, illustrating plasmon resonance behavior.

The combination of a high refractive index and strong reflectivity confirms its suitability for integration into photonic and energy-conversion systems where precise control of light propagation and reflection is essential. Furthermore, the slightly narrower band gap of the P63mc phase enhances absorption in the near-infrared spectral region, broadening its applicability for IR photodetectors, long-wavelength optoelectronic devices, and integrated photonic components. This contrast between the two phases highlights the potential of AlSb as a dual-range material capable of operating across both the visible and IR regimes.

The static optical constants summarized in Table 2 further clarify the phase-dependent optical response of AlSb. The cubic F-43m phase exhibits larger values of the real part of the dielectric constant $\varepsilon_1(0)$=12.21, refractive index n(0)=3.49, and reflectivity R(0)=0.31 compared to the hexagonal P6$_3$mc phase, which shows $\varepsilon1(0)$=9.30, n(0)=3.05, and R(0)=0.26. These differences reflect the stronger electronic polarizability of the cubic phase, consistent with its higher crystal symmetry and more effective *sp³* hybridization.

**Table 2.** Static optical parameters of cubic and hexagonal phases of AlSb, including the refractive index n(0), real part of the dielectric constant $\varepsilon_1(0)$, and reflectivity R(0).

| AlSb | $\varepsilon_1(0)$ | n(0) | R(0) |
|---|---|---|---|
| F43m | 12.208 | 3.49 | 0.308 |
| P6$_3$mc | 9.927 | 3.04 | 0.256 |

The thermoelectric properties of AlSb were analyzed to determine how crystal structure and temperature affect the Seebeck coefficient, electrical conductivity, and power factor. The Seebeck coefficient, which quantifies the voltage generated by a temperature gradient, is highly sensitive to the electronic states near the Fermi level [54-60]. The analysis was performed using temperature- and carrier-concentration-dependent transport coefficients, allowing a direct assessment of thermopower generation, heat transport, and power conversion efficiency. The next section explores how these effects are reflected in the macroscopic thermoelectric performance of AlSb.

Figure 9 presents the temperature dependence of the Seebeck coefficient for both AlSb phases. In the entire investigated temperature range (300-1000 K), both structures exhibit large negative Seebeck coefficients, indicating dominant *n*-type transport. The magnitude of the Seebeck coefficient decreases monotonically with increasing temperature, reflecting the progressive thermal activation of charge carriers and the reduction of energy-dependent asymmetry in the transport distribution function. At low temperatures, both phases show very large absolute Seebeck values (exceeding 700-800 µV/K), characteristic of intrinsic or weakly doped semiconductors. As temperature increases, the cubic phase maintains a systematically higher Seebeck coefficient compared to the hexagonal phase, which can be attributed to its wider band gap and more symmetric band-edge dispersion, resulting in a stronger energy filtering effect for charge carriers.

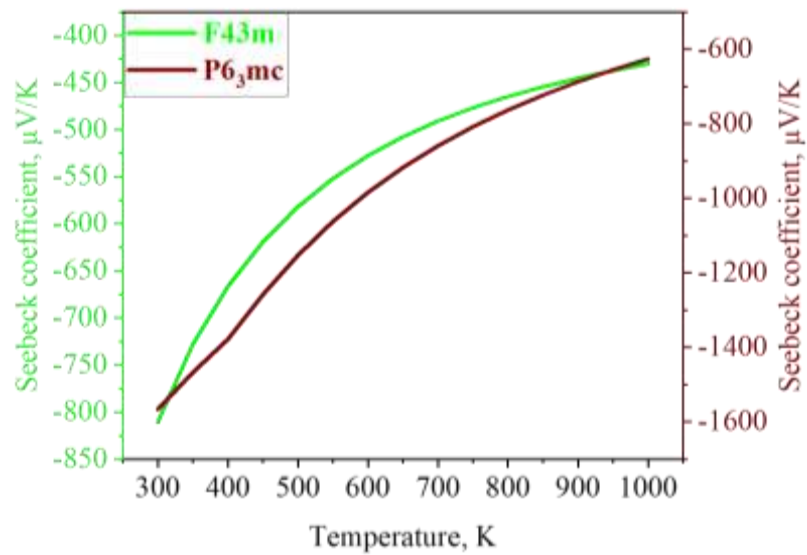

**Figure 9.** Temperature dependence of the Seebeck coefficient for cubic and hexagonal AlSb.

The temperature dependence of the carrier concentration is shown in Figure 10. In both phases, the carrier density increases rapidly with temperature, following the expected thermally activated behavior of narrow-gap semiconductors. The cubic phase exhibits a significantly higher carrier concentration at elevated temperatures, indicating more efficient carrier excitation and a higher density of accessible electronic states near the band edges. In contrast, the hexagonal phase shows a delayed carrier activation, consistent with its reduced symmetry and modified electronic structure.

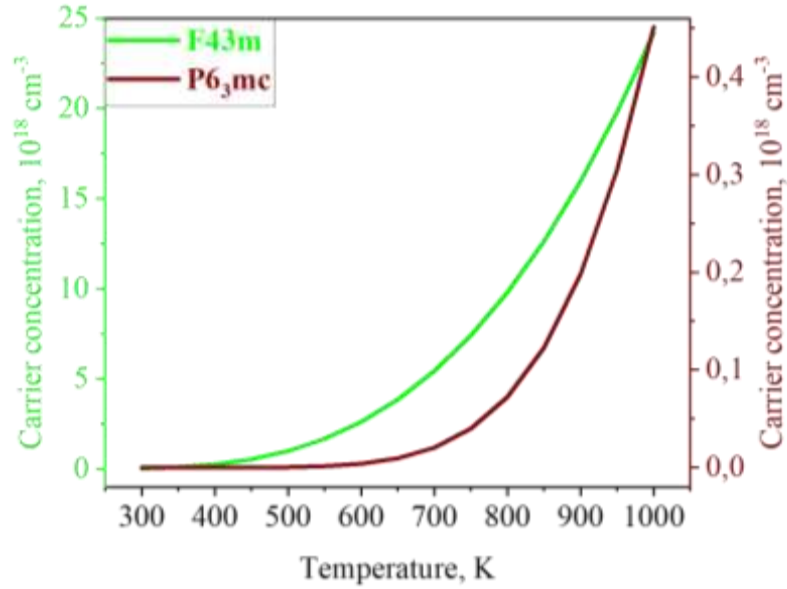

Figure 10. Temperature dependence of the carrier concentration for cubic and hexagonal AlSb

To further clarify the intrinsic transport behavior, the Seebeck coefficient was analyzed as a function of carrier concentration (Figure 11). In both phases, the magnitude of the Seebeck coefficient decreases with increasing carrier concentration, reflecting the transition from a low-density, energy-selective transport regime to a more metallic-like response. For a given carrier concentration, the cubic phase consistently exhibits a lower absolute Seebeck coefficient than the hexagonal phase, indicating a more delocalized electronic structure and reduced band-edge asymmetry.

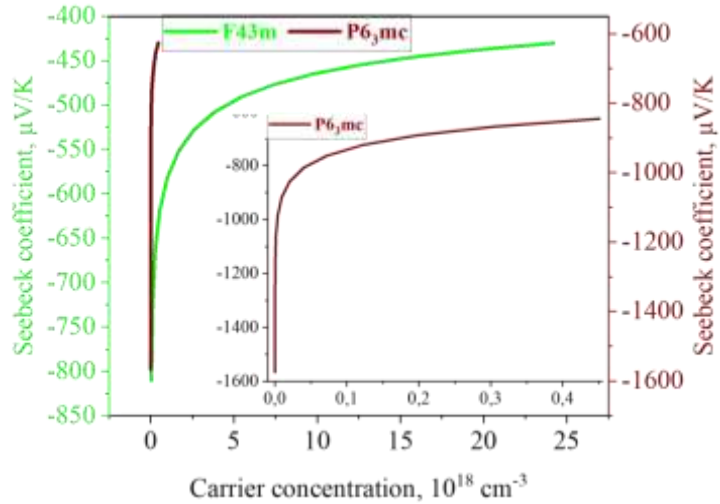

**Figure 11.** Seebeck coefficient as a function of carrier concentration for cubic and hexagonal AlSb.

The electronic thermal conductivity as a function of carrier concentration is shown in Figure 12. In both structures, thermal conductivity increases monotonically with carrier concentration, as enhanced carrier populations facilitate heat transport. However, the hexagonal $P6_3mc$ phase displays systematically lower thermal conductivity across the entire concentration range. This reduction can be attributed to its lower lattice symmetry and increased scattering channels, which suppress heat transport and are favorable for thermoelectric performance.

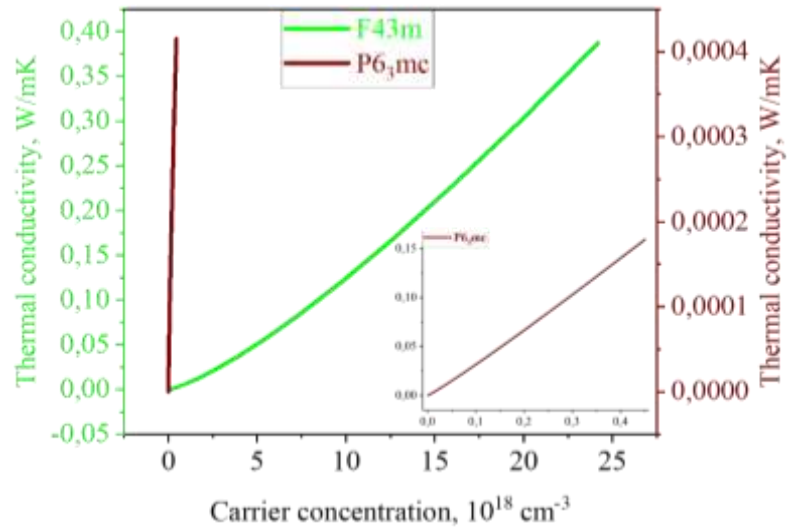

**Figure 12.** Thermal conductivity as a function of carrier concentration for cubic and hexagonal AlSb.

The combined effect of Seebeck coefficient and electrical transport is reflected in the power factor, shown in Figure 13. Both AlSb phases exhibit a monotonic increase in the power factor with carrier concentration, indicating improved thermoelectric efficiency under moderate to high doping conditions. The cubic F-43m phase reaches substantially higher power factor values, driven by its higher carrier concentration and stronger electrical conductivity. In contrast, the hexagonal phase shows a more gradual increase, limited by reduced carrier mobility but partially compensated by its lower thermal conductivity.

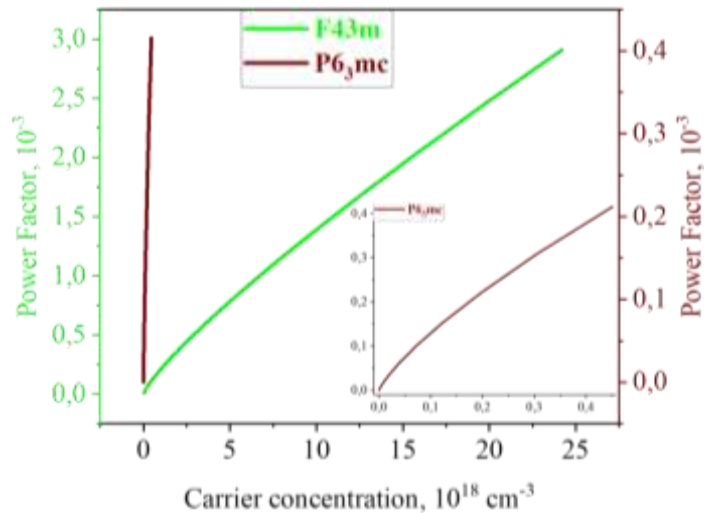

Figure 13. Power factor as a function of carrier concentration for cubic and hexagonal AlSb.

Tthe results demonstrate that the cubic phase is more favorable for applications requiring high power output and efficient charge transport, while the hexagonal phase offers advantages in regimes where reduced thermal conductivity and enhanced thermopower are critical. This complementary behavior highlights the potential of AlSb as a phase-tunable thermoelectric material, where structural modification provides a pathway to optimize performance across different operating conditions. Notably, the observed differences in thermoelectric response between the cubic and hexagonal phases are consistent with their temperature-dependent free-energy landscape, where the gradual stabilization of the P6$_3$mc phase at elevated temperatures, as indicated by ΔF(T), correlates with its reduced thermal conductivity and enhanced thermopower-driven transport behavior. The findings suggest that targeted structural tuning and phonon engineering in AlSb-based compounds can further optimize their high-temperature thermoelectric properties and expand their applicability in next-generation energy conversion and waste-heat recovery technologies [59-62].

## CONCLUSION

A comprehensive first-principles investigation of the cubic (F-43m) and hexagonal (P6$_3$mc) phases of AlSb has been performed to elucidate the role of crystal symmetry and bonding in shaping their electronic, optical, and thermoelectric properties. By combining SCAN-relaxed geometries with mBJ+U electronic structure calculations, the band-edge states are described with improved accuracy, explicitly accounting for the influence of Sb semicore *d* electrons, which is essential for reliable prediction of functional properties. Both polymorphs are identified as direct-gap semiconductors, with calculated band gaps of approximately 1.71 eV for the cubic phase and 1.50 eV for the hexagonal phase, consistent with available experimental data. The cubic structure exhibits stronger band dispersion and higher carrier mobility, resulting in enhanced optical response in the visible and near-infrared region, moderate reflectivity, and a larger static refractive index (≈3.5). These features favor optoelectronic operation under high electric fields and stable visible-range detection. In contrast, the hexagonal phase shows a reduced band gap, increased electronic-state density near the band edges, and lower thermal conductivity, which collectively enhance its thermoelectric performance at elevated temperatures. The calculated Seebeck coefficients reach several hundred μV/K over a wide temperature range, while the power factor increases monotonically with carrier concentration, indicating efficient thermoelectric conversion under moderate to high doping conditions. The lower reflectivity and refractive index further support enhanced infrared absorption in the hexagonal modification.

Across all calculated properties, a consistent structure-property relationship is observed: symmetry-driven differences in band dispersion and density of states control optical transition strength and charge transport, which are intrinsically coupled to thermal transport behavior. This interplay explains the superior optoelectronic performance of the cubic phase and the enhanced thermoelectric response of the hexagonal phase. The results demonstrate that AlSb is a versatile III-V semiconductor whose functional characteristics can be tuned through crystal-phase engineering. The present study provides a quantitative theoretical framework for designing AlSb-based materials and heterostructures aimed at integrated optoelectronic and thermoelectric applications, where simultaneous optimization of photon absorption, carrier transport, and heat-to-electric energy conversion is required.

**FUNDING**

This work was supported by the Interstate Fund for Humanitarian Cooperation of the CIS Member States through the scientific project funded by the International Nanotechnology Innovation Center of the CIS (grant no. 25-113), and by the International Science and Technology Center (ISTC) (grant no. TJ-0040).

**AUTHOR CONTRIBUTIONS**

DN and AB: Investigation, Methodology, Supervision, Writing - original draft, Writing - review & editing; ShM and IR: Investigation, Conceptualization, Data curation, Visualization; Funding acquisition; SM: Formal analysis, Validation; SS: Resources, Data curation; AA and SD: Visualization, Editing assistance; MBI and KhKh: Methodology, Writing - review & editing.

**CONFLICT OF INTEREST**.

The authors declare no conflicts of interest regarding the publication of this manuscript.

**USE OF AI TOOLS**

Artificial intelligence tools were used solely for language polishing, grammatical correction, and minor editorial assistance. All scientific content, data analysis, and interpretation of the results were performed exclusively by the authors. The authors carefully reviewed and verified all AI-assisted text to ensure accuracy, originality, and compliance with ethical publishing standards, including the COPE guidelines.